# On the determination of the trilinear boson couplings in $e^+e^- \to \ell\bar{\nu}_\ell\, q\bar{q}'$ at LEPII.

Costas G. Papadopoulos

Physics Department, Durham University, Durham, DH1 3LE, UK
and
CERN, Theory Division, CH-1211 Geneva 23, Switzerland

ABSTRACT

We present the full tree-order calculation of the process $e^+e^- \to \ell\bar{\nu}_\ell\, q\bar{q}'$ including the width effects in a way which is consistent with gauge invariance and high-energy unitarity and which has a natural interpretation within standard perturbation theory. We also include C and P preserving deviations from Standard Model trilinear vector boson couplings. An analysis of the sensitivity of these channels on the trilinear couplings, using well measured angular variables of the four-fermion final state, is also presented.

LEP II will provide us, for the first time, with the opportunity to study *directly* the $W$ boson, above the energy threshold for $W$-pair production. One of the most important measurements at these energies will be the determination of the trilinear vector boson couplings [1, 2, 3, 4, 5], a characteristic manifestation of the underlying non-Abelian symmetry of elementary particle interactions [6]. In order to determine these couplings we need accurate predictions of the cross sections which will be measured at LEP II. It is therefore evident that a good theoretical description of the four fermion production is indispensable.

In this paper we present a full tree-order calculation of the processes:

$$e^+e^- \to \ell\bar{\nu}_\ell \, q\bar{q}' \tag{1}$$

taking into account the non-zero width of the heavy unstable particles $W$ and $Z$ in a way which is consistent with gauge invariance and well justified within standard perturbation theory. We also include all possible deviations[1] from the Standard Model trilinear vector boson interactions which respect C and P invariance. The need for this calculation stems from two different directions: from the experimental point of view it is evident that the LEP II environment will consist of four-fermion final states; from the theoretical point of view, we expect that finite width effects as well as non-resonant contributions, estimated to be of the order of a few per cent, will be significant for the trilinear vector boson couplings measurements.

The Feynman diagrams contributing to the processes Eq.(1) are given in Figs.1,2. Graphs 1 to 3 contribute to $W$-pair production, whereas graphs 1 to 6 as well as 11 to 16 contribute to single-$W$ production, $e^+e^- \to e^-\bar{\nu}_e W^+$ [4, 5]. Graphs 1, 2, 11 and 12 are the signal graphs, since they include the trilinear boson vertex function, whereas graphs 1 to 3 are the double-resonant graphs, 4 to 16 are the single-resonant ones and finally graphs 17 to 20 are the non-resonant ones.

The amplitudes have been calculated using the $E$-vector formulation [4] of the spinor product technique [7]. This is based on the following representation for the fermion 'current', for massless fermions:

$$E^\mu_\lambda(p_1, p_2) \equiv \bar{u}_\lambda(p_1)\gamma^\mu u_\lambda(p_2) \tag{2}$$

where

$$E^0_- = \sqrt{p_1^+ p_2^+} + \frac{(p_{1x} + ip_{1y})(p_{2x} - ip_{2y})}{\sqrt{p_1^+ p_2^+}}$$

$$E^x_- = \sqrt{\frac{p_2^+}{p_1^+}}(p_{1x} + ip_{1y}) + \sqrt{\frac{p_1^+}{p_2^+}}(p_{2x} - ip_{2y})$$

$$E^y_- = -i\left(\sqrt{\frac{p_2^+}{p_1^+}}(p_{1x} + ip_{1y}) - \sqrt{\frac{p_1^+}{p_2^+}}(p_{2x} - ip_{2y})\right)$$

---

[1] Sometimes referred to as anomalous couplings.

$$E_-^z = \sqrt{p_1^+ p_2^+} - \frac{(p_{1x} + ip_{1y})(p_{2x} - ip_{2y})}{\sqrt{p_1^+ p_2^+}} \qquad (3)$$

with $p^\pm = p^0 \pm p^3$.

The main advantage of this formulation is that there is a very straightforward way to write down the expression for a given Feynman graph. For example, for the graph number 7 the expression is simply (omitting propagators and coupling constants):

$$T \sim \sum_{i=q,e^-,e^+} \text{sign}(p_i) E_-(p_q, p_i) \cdot E_\lambda(p_{e^+}, p_{e^-})\ E_-(p_i, p_{\bar{q}'}) \cdot E_-(p_\ell, p_{\nu_\ell}) \qquad (4)$$

where $\text{sign}(p) = \pm 1$ for outgoing and incoming momenta respectively. Furthermore the FORTRAN code representation of the amplitudes is very simple, facilitating considerably the writing and checking of the programme.

One of the first problems, inherent in any four-fermion calculation, is the treatment of the propagator of the heavy unstable particles, $W$ and $Z$, appearing as intermediate states in all Feynman graphs [8]. The problem is that if we implement the tree-order form of this propagator

$$\frac{1}{p^2 - m_V^2}$$

we have a non-integrable singularity in phase space, located at $p^2 = m_V^2$. The well-established procedure to cure this problem is to take into account the finite width of these particles. The propagator takes the well-known Breit-Wigner form

$$\frac{1}{p^2 - m_V^2 + im_V \Gamma_V(p^2)}.$$

From the point of view of the quantum field theory, the Breit-Wigner propagator can be understood as the effect of the resummation of all self-energy graphs as shown in Fig.3. The 'width' of the particle is then identified with the imaginary part of the one-loop self-energy contribution and it is a function of $p^2$. The 'fixed' width $\Gamma_V$ is now defined at $p^2 = m_V^2$, and it is connected to the decay of the on-shell heavy particle, as is dictated by the optical theorem.

Although all these seem to be rather elementary, the treatment of the width effects, using the above prescription, leads in general to gauge-invariance violation [9], as well as to unitarity violation at high energies. Let us first see how this problem appears in the 't-channel' graphs 11, 14, 17 and 19 of Fig.2, where a photon is exchanged carrying a space-like momentum. Their expressions, omitting trivial factors, are given as follows:

$$T_{11} = \frac{1}{(p_1 - p_3)^2}\ \frac{1}{(p_4 + p_{56})^2}\ \frac{1}{2p_5 \cdot p_6 - m_W^2}$$
$$\bar{u}_3 \gamma^\mu u_1\ \bar{u}_5 \gamma^\mu u_6\ \bar{u}_2 \gamma_\mu (\slashed{p}_4 + \slashed{p}_{56}) \gamma_\nu u_4$$
$$T_{14} = \frac{1}{(p_1 - p_3)^2}\ \frac{1}{-2p_2 \cdot p_4 - m_W^2}\ \frac{1}{2p_5 \cdot p_6 - m_W^2}$$

$$\mathcal{T}_{17} = \frac{\bar{u}_3\gamma^\mu u_1 \; \bar{u}_5\gamma^\alpha u_6 \; \bar{u}_2\gamma^\nu u_4 \;\; \Gamma^\gamma_{\mu\nu\alpha}}{(p_1-p_3)^2 \; (p_1-p_3-p_5)^2 \; -2p_2\cdot p_4 - m_W^2}$$

$$\mathcal{T}_{19} = \frac{\bar{u}_3\gamma^\mu u_1 \; \bar{u}_5\gamma_\mu(\not{p}_3+\not{p}_5-\not{p}_1)\gamma_\nu u_6 \; \bar{u}_2\gamma^\nu u_4}{(p_1-p_3)^2 \; (p_1-p_3-p_6)^2 \; -2p_2\cdot p_4 - m_W^2}$$
$$\bar{u}_3\gamma^\mu u_1 \; \bar{u}_5\gamma_\mu(\not{p}_1-\not{p}_3-\not{p}_6)\gamma_\nu u_6 \; \bar{u}_2\gamma^\nu u_4 \tag{5}$$

where $p_1, p_3, p_4, p_5, p_6$ are the momenta of the incoming electron and the outgoing electron, neutrino, quark and antiquark, respectively, and $\Gamma^\gamma_{\mu\nu\alpha}$ is the three-boson vertex function. The singularity appearing in the collinear limit $p_3\|p_1$ is integrable due to the non-zero electron mass and the result is nothing but the well-known logarithmic rise of the total cross section, $\log(s/m_e^2)$, where $m_e$ is the electron mass, which forms the basis of the Weiszäker-Williams approximation [10]. If we look more closely at the expressions given above, we see that in order to get a logarithmic behaviour (or a logarithmic divergence in the case of massless electrons) we need to have the following relation

$$(p_1-p_3)^2(\mathcal{T}_{11}+\mathcal{T}_{14}+\mathcal{T}_{17}+\mathcal{T}_{19}) = 0$$

in the limit $p_3^\mu = x \; p_1^\mu$, which is nothing but the gauge invariance at the subprocess level $\gamma + e^+ \to \bar{\nu}_e q \bar{q}'$. Let us now write the above equation in the collinear limit:

$$\begin{aligned} 0 =& \left(\frac{2p_1\cdot p_2}{(p_4+p_5+p_6)^2} - \frac{2p_1\cdot(p_5+p_6)}{2p_2\cdot p_4 + m_W^2 - i\delta}\right)\frac{E_{56}\cdot E_{24}}{(p_5+p_6)^2 - m_W^2 + i\gamma} \\ & - \frac{\delta\Gamma_{\mu\alpha\beta}E_{24}^\beta E_{56}^\alpha p_1^\mu}{(p_5+p_6)^2 - m_W^2 + i\gamma} + \frac{1}{1-x}\frac{E_{56}\cdot E_{24}}{2p_2\cdot p_4 + m_W^2 - i\delta} \end{aligned} \tag{6}$$

where we have allowed for several possible contributions, including the width of the $W$ both in time-like and space-like regions ($\gamma$ and $\delta$ respectively) as well as a modification of the three-boson vertex. The solutions are easily worked out and they can be enumerated as follows ($\delta\Gamma_{\mu\alpha\beta}E_{24}^\beta E_{56}^\alpha p_1^\mu = \delta\Gamma \; E_{56}\cdot E_{24}/(1-x)$):

1. $\delta = \gamma = \delta\Gamma = 0$ which is the standard tree-order result. This of course has the disadvantage that the matrix element squared is not integrable over the phase space. A 'solution' to this problem, first proposed in reference [11], is to multiply the tree-order amplitude by a factor

$$\frac{q^2 - m_W^2}{q^2 - m_W^2 + im_W\Gamma_W}$$

where $q^2$ is the (positive) momentum squared of the intermediate $W$. This of course provides a gauge-invariant quantity, as an approximation of the tree-order cross section, which, nevertheless, is not well justified within standard perturbation theory.

2. $\delta = \gamma = m_W \Gamma_W$ and $\delta\Gamma = 0$ which is the so-called 'naive scheme', where a fixed width is considered in any $W$ propagator. This 'solution' shares the same problem as the first one, since there is no justification of the width-term in propagator functions with space-like momentum flow, where the imaginary part of the self-energy contribution is vanishing.

3. Finally, the last solution is given by:

$$\gamma = \delta\Gamma = \frac{\Gamma_W}{m_W}(p_5 + p_6)^2.$$

It is exactly this last solution which is the basis of our approach to the width problem. What it suggests is that we have to take into account the relevant contribution from the one-loop correction to the three-boson vertex [12]. This is shown in Fig.4, and we will use the term Breit-Wigner triangle, in order to emphasize the relation of this correction to the resummed Breit-Wigner propagator.

It is very easy to verify that the imaginary part of the one-loop graph, shown in Fig.5, fits perfectly the requirement of our third solution. It is not difficult to realize that the relation derived in the collinear limit is nothing but the standard Ward identity, which relates the imaginary part of the one-loop vertex with the inverse of the Breit-Wigner propagator[2].

Let us now come to the high-energy unitarity, which is related to the double-resonant graphs 1, 2 and 3. In the high energy limit the dominant part of each amplitude can be obtained by the replacements:

$$\bar{u}_5 \gamma^\alpha u_6 \to p_5^\alpha + p_6^\alpha \quad \text{and} \quad \bar{u}_3 \gamma^\beta u_4 \to p_3^\beta + p_4^\beta.$$

Then, all amplitudes simplify a lot and become proportional to

$$s\, E_{21} \cdot (p_5 + p_6 - p_3 - p_4),$$

where $s = (p_1 + p_2)^2$. The cancellation of these terms, whose contribution to the cross section grows linearly with $s$, is guaranteed by the following equations ($g_f^{(\lambda)}$ is the coupling of $Z$ with the fermion $f$ and $g_{ZWW}$ is the coupling with the $W$ boson):

$$Q_e + g_e^{(+)} g_{ZWW} = 0 \quad \text{and} \quad Q_e + g_e^{(-)} g_{ZWW} + \frac{1}{2s_w^2} = 0$$

a consequence of the $SU(2) \times U(1)$ gauge symmetry. It is more convenient to rewrite them in the following form

$$\vec{e}_\lambda \cdot \vec{g}_W + \delta_{\lambda,-} \frac{1}{2s_w^2} = 0$$

---

[2] For a comprehensive analysis on the treatment of the width of unstable particles in perturbation theory, see reference [13].

where
$$\vec{e}_\lambda = \begin{pmatrix} Q_e \\ g_e^{(\lambda)} \end{pmatrix}$$
and
$$\vec{g}_W = \begin{pmatrix} 1 \\ g_{ZWW} \end{pmatrix}$$

When we consider the width effects we have to resum the self-energy contributions for the neutral boson propagator. The building block of this resummation is, as usual, the imaginary part of the one-loop contribution, which can be written as

$$\mathcal{K} = -i\,e^2\,\frac{1}{24\pi}\,N_g \begin{pmatrix} 2\sum_f Q_f^2 & \sum_f Q_f(g_L^f + g_R^f) \\ \sum_f Q_f(g_L^f + g_R^f) & \sum_f ((g_L^f)^2 + (g_R^f)^2) \end{pmatrix}$$

where $N_g$ is the number of fermion generations, and the summation over the fermions, $\sum_f$, is restricted within a complete generation. In deriving this formula we have neglected all fermion masses, which are irrelevant in the high energy limit. It is a matter of algebra to realize that the vector $\vec{g}_W$ is an eigenvector of the matrix $\mathcal{K}$ and therefore the resummed result can be written in the form

$$(Q_e + g_e^{(+)} g_{ZWW})(1 + i\,e^2\,\frac{1}{24\pi}\,N_g \frac{2}{s_w^2})^{-1}$$

and
$$(Q_e + g_e^{(-)} g_{ZWW})(1 + i\,e^2\,\frac{1}{24\pi}\,N_g \frac{2}{s_w^2})^{-1} + \frac{1}{2s_w^2}$$

for right- and left-handed electrons respectively. As is easily seen, the right-handed amplitude automatically satisfies the high-energy unitarity condition, since it is proportional to the tree order result. For the left-handed part we have to add the relevant Breit-Wigner triangle contribution. There are three cut graphs contributing in this kinematical configuration, but only the cut on the neutral boson site is relevant for the high-energy limit. In this limit the effect of the triangle graph is given by the tree-order result multiplied by the factor $(1 + i\,e^2\,\frac{1}{24\pi}\,N_g \frac{2}{s_w^2})$, which cancels exactly the self-energy resummed contribution. This result also relies on the gauge symmetry relations:

$$\sum_f{}' Q_f = 4 N_g \quad \text{and} \quad \sum_f{}' g_f^{(-)} = 4 N_g g_{ZWW}$$

where $\sum_f{}'$ means that $t = 1/2$ and $t = -1/2$ fermions are taken with a relative minus sign, which gives rise to a complete factorization of the Breit-Wigner triangle graph contribution, in the high energy limit.

Finally we would like to briefly comment on the reason why the sum of the Breit-Wigner propagator and the Breit-Wigner triangle form a subset of gauge-invariant graphs. As far as the high-energy limit is concerned, the result is proportional to $N_g \alpha$, where $\alpha = e^2/4\pi$ is the electromagnetic coupling constant, since the loop summation is taken over complete fermion generations. It is rather elementary to realize that in the limit $N_g \to \infty$ and $\alpha \to 0$, with $N_g \alpha$=fixed, the only graphs that survive are exactly those considered in our discussion and therefore they are by themselves gauge invariant.

It is worth while to notice that, at the amplitude level, the cancellation of the most divergent term, which is proportional to $s/m_W^2$, holds for each fermion separately. The reason is that the triangle contribution is proportional to the vector

$$\vec{g}'_W = \frac{1}{s_W^2} \begin{pmatrix} t_f Q_f \\ t_f(t_f - Q_f s_w^2)\frac{1}{s_w c_w} \end{pmatrix}$$

whereas the corresponding $\mathcal{K}$-matrix is proportional to

$$\mathcal{M} = \begin{pmatrix} 2Q_f^2 & Q_f(t_f - 2Q_f s_w^2)\frac{1}{s_w c_w} \\ Q_f(t_f - 2Q_f s_w^2)\frac{1}{s_w c_w} & (t_f^2 - 2t_f Q_f s_w^2 + 2Q_f^2 s_w^4)\frac{1}{s_w^2 c_w^2} \end{pmatrix}$$

with the property that

$$\mathcal{M} \vec{g}_W = -\vec{g}'_W \ .$$

Therefore the triangle contribution cancels exactly the resummed self-energy one, and terms growing linearly with $s$ are absent, as unitarity requires.

The bosonic contribution to the width effects should be treated more carefully. First of all the bosonic self-energy contribution can be resummed in the same way as the fermionic one and it is easily proved that the vector $\vec{g}_W$ is still an eigenvector of the corresponding $\mathcal{K}$-matrix, in the high energy limit. Nevertheless, including the triangle graph contribution does not, generically, lead to a gauge invariant result and one has to take into account box graphs as well [13, 14]. Furthermore it is not possible to isolate a part of the higher order corrections, which is gauge invariant, in a way similar to the one suggested by the high $N_g$-limit in the fermionic sector. A possible alternative is to use background field techniques[15] in order to restore the gauge invariance. It should be emphasized, however, that, due to threshold effects arising from the phase-space integration, the bosonic graphs do not contribute to the definition of the fixed width of the $W$ and $Z$, at this order of perturbation theory. The same is still true for the $U(1)$ gauge-invariance restoration in the small momentum transferred limit, $q_\gamma^2 \to 0$. Finally their contribution at LEP II energies is supressed by factors of the form $1 - 4m_V^2/s$, where $m_V$ refers to the mass of $W$, $Z$ and Higgs particle.

After completing the amplitude calculation we now proceed with the integration over the phase space. To this end we have implemented a Monte Carlo integration algorithm which is essentially identical to the multichannel approach of references [16, 17]. The problem is that the amplitude we have to integrate over is a very complicated function of the kinematical variables, peaking at different regions of phase space. The idea is to define different kinematical mappings, corresponding to different peaking structures of the amplitude and then use an optimization procedure to adjust the percentage of the generated phase-space points, according to any specific mapping, in such a way that the total error is minimized. Notice that the four particle phase space is determined by eight independent variables. For each algorithm, the set of these variables and the corresponding mapping to the eight pseudorandom numbers, $\rho_i$, $i = 1, \ldots, 8$, is given below:

$$
\begin{aligned}
m_{q,\bar{q}'}^2 &= m_W^2 + m_W \Gamma_W \tan(\rho_1(y_+ - y_-) + y_-) \\
y_+ &= \tan^{-1}(\frac{E^2 - m_W^2}{m_W \Gamma_W}) \\
y_- &= \tan^{-1}(-\frac{m_W}{\Gamma_W}) \\
m_{\ell,\bar{\nu}_\ell}^2 &= m_W^2 + m_W \Gamma_W \tan(\rho_2(y'_+ - y_-) + y_-) \\
y'_+ &= \tan^{-1}(\frac{(E - m_{q,\bar{q}'})^2 - m_W^2}{m_W \Gamma_W}) \\
\cos\theta_W &= 2\rho_3 - 1, \quad \phi_W = 2\pi\rho_4 \\
\cos\theta_\ell^* &= 2\rho_5 - 1, \quad \phi_\ell^* = 2\pi\rho_6 \quad \text{(in the } \ell, \bar{\nu}_\ell \text{ rest frame)} \\
\cos\theta_q^* &= 2\rho_7 - 1, \quad \phi_q^* = 2\pi\rho_8 \quad \text{(in the } q, \bar{q}' \text{ rest frame)} 
\end{aligned} \quad (7)
$$

which simulates the peaking structure of graphs 1 and 2,

$$
\begin{aligned}
m_{q,\bar{q}'}^2 &= m_W^2 + m_W \Gamma_W \tan(\rho_1(y_+ - y_-) + y_-) \\
m_{q,\bar{q}',\bar{\nu}_\ell}^2 &= (\rho_2 E^{2x} + (1 - \rho_2) m_{q,\bar{q}'}^{2x})^{1/x} \\
\cos\theta_\ell &= 1 - \left((1 - \rho_3)(1 - c_{min})^{(1-\nu)} + \rho_3(1 - c_{max})^{(1-\nu)}\right)^{1/(1-\nu)} \\
\phi_\ell &= 2\pi\rho_4 \\
\cos\theta_{\bar{\nu}_\ell}^* &= 2\rho_5 - 1, \quad \phi_{\bar{\nu}_\ell}^* = 2\pi\rho_6 \quad \text{(in the } \bar{\nu}_\ell, q, \bar{q}' \text{ rest frame)} \\
\cos\theta_q^* &= 2\rho_7 - 1, \quad \phi_q^* = 2\pi\rho_8 \quad \text{(in the } q, \bar{q}' \text{ rest frame)}
\end{aligned} \quad (8)
$$

which for $x = 1$ simulates the peaking structure of graph 11 and with $x \to 0$ the peaking structure of graphs 14, and finally

$$
\begin{aligned}
m_{q,\bar{q}'}^2 &= m_W^2 + m_W \Gamma_W \tan(\rho_1(y_+ - y_-) + y_-) \\
m_{\ell,\bar{\nu}_\ell}^2 &= m_W^2 + m_W \Gamma_W \tan(\rho_2(y'_+ - y_-) + y_-) \\
\cos\theta_W &= \alpha - \left((1 - \rho_3)(\alpha - 1)^{(1-\nu)} + \rho_3(\alpha + 1)^{(1-\nu)}\right)^{1/(1-\nu)} \\
\alpha &= \frac{E p_W^0 - m_{\ell,\bar{\nu}_\ell}^2}{E|\vec{p}_W|} \\
\phi_W &= 2\pi\rho_4 \\
\cos\theta_\ell^* &= 2\rho_5 - 1, \quad \phi_\ell^* = 2\pi\rho_6 \quad \text{(in the } \ell, \bar{\nu}_\ell \text{ rest frame)} \\
\cos\theta_q^* &= 2\rho_7 - 1, \quad \phi_q^* = 2\pi\rho_8 \quad \text{(in the } q, \bar{q}' \text{ rest frame)}
\end{aligned} \quad (9)
$$

which simulates the peaking structure of graph 3.

In order to study the sensitivity of the processes under consideration on the trilinear boson couplings we need a parametrization of these interactions that goes beyond the Standard Model . There are of course several possibilities, but we are going to restrict ourselves to the most economical one. The relevant interaction Lagrangian can be written in the following form [2]:

$$\mathcal{L} = \sum_{V=\gamma,Z} \left\{ g_V (V_\mu W^{-\mu\nu} W^{+\nu} - V_\mu W^{+\mu\nu} W^{-\nu} + \kappa_V V_{\mu\nu} W^{+\mu} W^{-\mu\nu}) \right.$$
$$\left. + g_V \frac{\lambda_V}{M_W^2} V_{\mu\rho} W^{+\rho\nu} W_\nu^{-\mu} \right\} \tag{10}$$

where

$$W_{\mu\nu}^\pm = \partial_\mu W_\nu^\pm - \partial_\nu W_\mu^\pm$$

$W^\pm$ is the $W$-boson field, and $g_\gamma = e$, $g_Z = e\,\text{ctg}\,\theta_w$, $\kappa_\gamma = \kappa_Z = 1$ and $\lambda_\gamma = \lambda_Z = 0$ at tree order in the Standard Model . It is more convenient to express the different couplings in terms of their deviations from the Standard Model values. For this we define the following deviation parameters:

$$\delta_Z = g_Z - \text{ctg}\,\theta_w$$
$$x_\gamma = \kappa_\gamma - 1$$
$$x_Z = (\kappa_Z - 1)(\text{ctg}\,\theta_w + \delta_Z) \tag{11}$$

It is worth while to note that the interaction Lagrangian becomes linear with respect to the above parameters (including also $\lambda_\gamma$ and $\lambda_Z$).

The $VW^+W^-$ vertex function can be derived from the above interaction Lagrangian and has the following form [4]:

$$\Gamma^V_{\mu\alpha\beta}(q, p_+, p_-) = ig_V \left\{ g_{\alpha\mu} \left[ (\kappa + \lambda \frac{p_-^2}{m_W^2})q - p_+ \right]_\beta + \left( g_{\alpha\beta} - \lambda \frac{q_\alpha q_\beta}{m_W^2} \right)(p_+ - p_-)_\mu \right.$$
$$+ g_{\beta\mu} \left[ p_- - (\kappa + \lambda \frac{p_+^2}{m_W^2})q \right]_\alpha - \frac{\lambda}{m_W^2} q^2 g_{\alpha\beta} p_{-\mu}$$
$$+ \frac{\lambda}{m_W^2} p_{+\alpha}(g_{\mu\beta} p_+ \cdot q - p_{+\mu} q_\beta) - \frac{\lambda}{m_W^2} p_{-\beta}(g_{\mu\alpha} p_- \cdot q - p_{-\mu} q_\alpha)$$
$$\left. + \frac{\lambda}{m_W^2} p_- \cdot q \; q_\mu g_{\alpha\beta} \right\} \tag{12}$$

where $p_\pm$ are the momenta of $W^\pm$ respectively, and $q$ the momentum of the neutral gauge boson ($\gamma$, $Z$). Note that in many processes a great simplification occurs due to fermion current conservation; for example the term proportional to $q_\mu$ vanishes whenever the neutral boson is coupled to a massless fermion pair.

During the last few years, considerable progress has been achieved concerning the understanding of the physics underlying the non-standard boson self-couplings. As Gounaris and Renard [18] showed, the deviations from the Standard Model couplings can be parametrized in a manifestly gauge-invariant (but still non-renormalizable) way, by considering gauge-invariant operators involving higher-dimensional interactions among gauge bosons and Higgs field. Restricting ourselves to $SU(2)_L \times U(1)_Y$-invariant operators with dimension up to 8, we can have the following list [5, 19]:

$$\begin{aligned}
\mathcal{O}_{B\Phi} &= B^{\mu\nu}(D_\mu\Phi)^\dagger(D_\nu\Phi) \\
\mathcal{O}_{W\Phi} &= (D_\mu\Phi)^\dagger \boldsymbol{\tau} \cdot \boldsymbol{W}^{\mu\nu}(D_\nu\Phi) \\
\mathcal{O}_W &= \frac{1}{3!}(\boldsymbol{W}^\mu_{\ \rho} \times \boldsymbol{W}^\rho_{\ \nu}) \cdot \boldsymbol{W}^\nu_{\ \mu} \\
\mathcal{O}'_{W\Phi} &= (\Phi^\dagger \boldsymbol{\tau} \cdot \boldsymbol{W}^{\mu\nu}\Phi)(D_\mu\Phi)^\dagger(D_\nu\Phi) \\
\mathcal{O}'_W &= (\boldsymbol{W}^\mu_{\ \rho} \times \boldsymbol{W}^\rho_{\ \nu}) \cdot (\Phi^\dagger \frac{\boldsymbol{\tau}}{2}\Phi)B^\nu_{\ \mu}
\end{aligned} \quad (13)$$

where $\tau_i = \frac{1}{2}\sigma_i$ ($\sigma_i$ are the Pauli matrices),

$$B_{\mu\nu} = \partial_\mu B_\nu - \partial_\nu B_\mu$$

where $B_\mu$ is the $U(1)_Y$ gauge field,

$$\boldsymbol{W}_{\mu\nu} = \partial_\mu \boldsymbol{W}_\nu - \partial_\nu \boldsymbol{W}_\mu - g_2 \boldsymbol{W}_\mu \times \boldsymbol{W}_\nu$$

where $\boldsymbol{W}$ are the $SU(2)_L$ gauge fields and

$$\Phi = \begin{pmatrix} \phi^+ \\ \frac{1}{\sqrt{2}}(v + H + i\phi^0) \end{pmatrix}$$

is the Higgs doublet. The covariant derivative $D_\mu$ is given, as usual, by

$$D_\mu = \partial_\mu + i\, g_2 \boldsymbol{\tau} \cdot \boldsymbol{W}_\mu - i\, g_1 B_\mu$$

and $e = g_2 \sin\theta_w = g_1 \cos\theta_w$.

The interaction Lagrangian can be written now as

$$\mathcal{L} = \mathcal{L}_{SM} + \sum g_i \mathcal{O}_i \quad (14)$$

where $\mathcal{O}_i$ are the operators given in Eq.(13). The effect of each operator on the trilinear coupling parametrization, given in Eq.(10), is described in table 1. The main consequence of the above considerations is that $SU(2)_L \times U(1)_Y$ gauge invariance implies relations between deviation parameters, which must be taken into account in a phenomenological analysis. Furthermore, taking into account the limitting statistics expected at LEP II this is a very welcome result, since an investigation of the full multiparameter space seems to be rather impossible. In this paper, adopting this point of view, we focus our analysis on the six one-parameter cases, as given in table 1.

| case | operator | non-vanishing deviations |
|------|----------|--------------------------|
| 1 | $\mathcal{O}_{B\Phi}$ | $x_\gamma$, $x_Z = -x_\gamma \text{tg}\, \theta_w$ |
| 2 | $\mathcal{O}_{W\Phi} - \mathcal{O}_{B\Phi}$ | $\delta_Z$ |
| 3 | $\mathcal{O}_{W\Phi}$ | $x_\gamma$, $\delta_Z = x_\gamma/(s_w c_w)$, $x_Z = -x_\gamma \text{tg}\, \theta_w$ |
| 4 | $\mathcal{O}_W$ | $\lambda_\gamma$, $\lambda_Z = \lambda_\gamma$ |
| 5 | $\mathcal{O}'_{W\Phi}$ | $x_\gamma$, $x_Z = x_\gamma \text{ctg}\, \theta_w$ |
| 6 | $\mathcal{O}'_W$ | $\lambda_\gamma$, $\lambda_Z = -\lambda_\gamma \text{tg}^2 \theta_w$ |

Table 1: Contribution of different gauge-invariant operators to the anomalous couplings ($s_w = \sin \theta_w$ and $c_w = \cos \theta_w$).

Having clarified the amplitude as well as the phase-space calculation we now proceed with the presentation of our numerical results. Several checks have been made in order to guarantee their correctness. First of all both the $U(1)$ gauge invariance and the unitary high-energy behaviour of the calculation have been successfully tested at the numerical level. Furthermore we have checked our programme with the EXCALIBUR Monte Carlo [16], and taking identical conditions (width implementation, input parameters, kinematical cuts, etc) we found complete agreement at all energies at which we run the programme (175, 190, 200 and 500 GeV). It should be mentioned however that in what follows we have implemented our novel approach to the width effects, which is different from that of EXCALIBUR. Finally, since this calculation is an expansion of the single-$W$ production programme [4], several checks already done for this process are still in order for the four fermion calculation.

The input Standard Model parameters we have used are as follows:

$$M_W = 80.5 \text{ GeV}, \quad M_Z = 91.19 \text{ GeV}, \quad \sin^2 \theta_W = 0.23, \quad \alpha = 1/129$$

whereas the phase-space cuts are defined as:

$$\cos \theta_\ell \leq 0.95 \quad \cos \theta_{jet} \leq 0.95 \quad \cos \theta_{q,\bar{q}'} \leq 0.95$$
$$E_\ell, E_{jet} \geq 5 \text{ GeV} \quad \text{and} \quad m_{q,\bar{q}'} \geq 10 \text{ GeV} \qquad (15)$$

The 'width' of the $W$ is calculated by the formula:

$$\frac{\Gamma_W}{m_W} = \frac{\alpha}{12 s_w^2} n_f$$

where $n_f = 9(12)$ for $\sqrt{s} \leq m_{top} + m_b (\sqrt{s} > m_{top} + m_b)$. The $\gamma - Z$ propagator is defined by

$$\mathcal{D} = \frac{1}{s} \mathcal{D}_0 (1 - \mathcal{K} \mathcal{D}_0)^{-1}$$

where

$$\mathcal{D}_0 = \begin{pmatrix} 1 & 0 \\ 0 & \frac{s}{s - m_Z^2} \end{pmatrix}$$

As we have already pointed out, for the analysis we are attempting in this paper we will restrict ourselves to the one-parameter cases described in table 1. This simplifies the analysis, since the results (total cross sections and angular distributions) can always be written in the following simple form:

$$\sigma_i = \sigma_i^0 + x\sigma_i^i + x^2\sigma_i^s \tag{16}$$

where $x$ stands for any deviation parameter ($\delta_Z$, $x_\gamma$ or $\lambda_\gamma$), $\sigma_i^0$ is the Standard Model result, $\sigma_i^i$ is the interference term and $\sigma_i^s$ is the signal term squared. The subscript $i$ refers to a given phase-space configuration, or to a given phase-space variable. We find it more convenient to work with this parametrization rather than to specify some particular (arbitrary) values for the deviation parameters. In any case, the values of $\sigma_i^0$, $\sigma_i^i$ and $\sigma_i^s$ completely determine the dependence on $x$.

In Fig.6 we show the total cross section as a function of the energy from 175 GeV up to 500 GeV. The Standard Model cross section reaches its maximum around 200 GeV, whereas the interference term is negative and small in absolute value in the full energy range up to 500 GeV. The main contribution, especially at higher energies comes from $\sigma^s$, and it is more than an order of magnitude larger that the Standard Model prediction at 500 GeV. It is worth while to mention the difference in the energy growth between cases 3 and 4: this is to be expected, since the relations among the different deviation parameters given in case 3 are expected to result in a cancellation of the most divergent part of the amplitude[2].

In order to further analyse the sensitivity of the cross section to the trilinear couplings, we have to use differential distributions. There are in general two different and in some way complementary strategies in this analysis. The 'optimistic' one is to use the fully differential cross section, a function of the eight independent phase-space variables [2, 20]. This is of course the 'ultimate' information one can get, but since experimentally some of the phase-pace variables are poorly measured, one has to cope with a low efficiency factor. The 'realistic' one is to focus on these kinematical variables and the corresponding differential distributions, which are well measured experimentally and try to maximize the sensitivity with respect to the trilinear couplings. In this paper we are following the latter approach and we calculate the differential distribution with respect to the angle of the produced lepton in the laboratory frame. In order to compare the sensitivity of the lepton angle with the sensitivity of other variables, which are a priori more sensitive to the trilinear boson couplings, we calculate also the following differential distributions:

1. $d\sigma/d\cos\theta_W$, where $\theta_W$ is defined as the angle of the vector $\vec{p}_{q\bar{q}'}$ with the electron beam in the laboratory frame, which for the signal graphs is identical to the angle of $W^+$, and

2. $d\sigma/d\cos\theta_j^*$, where the angle $\theta_j^*$ is defined as the angle between the spatial momentum of the outgoing quark in the rest frame of the $q,\bar{q}'$ system, and the momentum $\vec{p}_{q\bar{q}'}$ in the laboratory frame, which for on-shell $W$ decay has been used as a polarization analyser [2].

The Standard Model distributions are given in Fig.7 for $\sqrt{s} = 175$ GeV and 205 GeV. As is evident from this figure, the distribution of the electron is very similar to the distribution of the $W^-$ (which is taken from the $W^+$ distribution by the replacement $\cos\theta \to -\cos\theta$) and peaks in the direction parallel to the electron beam, as a combined effect of the neutrino as well as photon t-channel exchange contributions.

The effect of the deviation terms on these distributions can be traced by studying the ratios

$$\frac{\sigma^i}{\sigma^0} \text{ and } \frac{\sigma^s}{\sigma^0}$$

as functions of the corresponding angular variables. As an example we present, in Fig.8, the above-mentioned quantities at $\sqrt{s} = 205$ GeV for the electron channel. Similar results are obtained at $\sqrt{s} = 175$ GeV as well as for the 'muon' channel, e.g. when a lepton different from the electron is produced in the final state. The main conclusion is that the angular dependence of the deviation terms is rather important and the general tendency is to flatten out the background distributions in all cases.

| Operator | $\theta_\ell$ | $\theta_j^*$ | $\theta_W$ |
|---|---|---|---|
| $\mathcal{O}_{B\Phi}$ | $-.37$ $.83$ | $-.62$ $1.13$ | $-.46$ $1.06$ |
| $x_\gamma$ | $-.17$ $.35$ | $-.30$ $.81$ | $-.21$ $.86$ |
| $\mathcal{O}_{W\Phi} - \mathcal{O}_{B\Phi}$ | $-.67$ $.62$ | $-.70$ $.70$ | $-.22$ $.24$ |
| $\delta_Z$ | $-.33$ $.41$ | $-.37$ $.55$ | $-.17$ $.26$ |
| $\mathcal{O}_{W\Phi}$ | $-.24$ $.31$ | $-.26$ $.33$ | $-.08$ $.09$ |
| $x_\gamma$ | $-.11$ $.28$ | $-.13$ $.28$ | $-.06$ $.07$ |
| $\mathcal{O}_W$ | $-.38$ $.40$ | $-.45$ $.46$ | $-.14$ $.16$ |
| $\lambda_\gamma$ | $-.16$ $.27$ | $-.21$ $.29$ | $-.09$ $.13$ |
| $\mathcal{O}'_{W\Phi}$ | $-.31$ $.41$ | $-.39$ $.60$ | $-.14$ $.15$ |
| $x_\gamma$ | $-.13$ $.22$ | $-.18$ $.50$ | $-.09$ $.11$ |
| $\mathcal{O}'_W$ | $-.44$ $.91$ | $-.62$ $.94$ | $-.42$ $.76$ |
| $\lambda_\gamma$ | $-.20$ $.53$ | $-.31$ $.53$ | $-.21$ $.50$ |

Table 2: Limits on deviation parameters from electron channel. The first row corresponds to $\sqrt{s} = 175$GeV and the second one to $\sqrt{s} = 205$GeV.

In order to determine the sensitivity to the trilinear vector boson couplings we implement a standard, though idealized, $\chi^2$ analysis. More precisely we define a '$\chi^2$' function as follows:

$$\chi^2 = \sum_i \frac{(\sigma_i - \sigma_i^0)^2}{\varepsilon_i^2}$$

| Operator | $\theta_\ell$ | $\theta_j^*$ | $\theta_W$ |
|---|---|---|---|
| $\mathcal{O}_{B\Phi}$ | $-.29$ $1.49$ | $-.45$ $1.60$ | $-.28$ $.38$ |
| $x_\gamma$ | $-.13$ $.19$ | $-.20$ $.95$ | $-.13$ $.17$ |
| $\mathcal{O}_{W\Phi} - \mathcal{O}_{B\Phi}$ | $-.51$ $.56$ | $-.56$ $.65$ | $-.15$ $.16$ |
| $\delta_Z$ | $-.26$ $.38$ | $-.31$ $.46$ | $-.11$ $.14$ |
| $\mathcal{O}_{W\Phi}$ | $-.17$ $.30$ | $-.20$ $.35$ | $-.05$ $.06$ |
| $x_\gamma$ | $-.08$ $.17$ | $-.10$ $.28$ | $-.04$ $.04$ |
| $\mathcal{O}_W$ | $-.24$ $.36$ | $-.33$ $.46$ | $-.09$ $.10$ |
| $\lambda_\gamma$ | $-.12$ $.21$ | $-.16$ $.26$ | $-.06$ $.07$ |
| $\mathcal{O}'_{W\Phi}$ | $-.23$ $.35$ | $-.29$ $.72$ | $-.09$ $.09$ |
| $x_\gamma$ | $-.09$ $.13$ | $-.13$ $.51$ | $-.05$ $.06$ |
| $\mathcal{O}'_W$ | $-.31$ $.98$ | $-.48$ $1.00$ | $-.30$ $.50$ |
| $\lambda_\gamma$ | $-.15$ $.51$ | $-.23$ $.52$ | $-.15$ $.32$ |

Table 3: Limits on deviation parameters from 'muon' channel. The first row corresponds to $\sqrt{s} = 175$GeV and the second one to $\sqrt{s} = 205$GeV.

where $\varepsilon_i$ is the statistical error of the $i$-th bin. In the present analysis we have chosen six equidistant bins in the range $-0.95 \leq \cos\theta \leq 0.95$, where $\theta$ stands for any of the angles under consideration. Furthermore, the integrated luminosity at $\sqrt{s} = 175$ GeV is taken to be $L = 500\text{pb}^{-1}$, whereas at $\sqrt{s} = 205$ GeV we have used the value $L = 300\text{pb}^{-1}$. The errors on the deviation parameters are now given by standard statistical theory [21] and the two standard deviation results are presented in tables ,. It should be mentioned, however, that the errors given in the above tables are in some sense a lower limit on the corresponding 'real' ones, since they do not take into account the goodness of the fit statistics, which relies on how well the real data are described by the Standard Model predictions. Nevertheless assuming that the real data would be described by the Standard Model and that $\chi^2/\text{dof} \sim 1$ we have a rather good estimate of the expected sensitivity. The main point underlined by this analysis is that, first of all, in most cases, the $\theta_W$ is more sensitive than the other variables, whereas $\theta_e$ is more efficient than $\theta_j^*$. Secondly, the sensitivity increases with increasing energy, even though the luminosity is substantially decreasing. Finally, electron and 'muon' channels seem to give comparable results, even though in the 'muon' channel we have doubled the corresponding statistics by including both $\mu$ and $\tau$ events, which share the same theoretical description. The limits obtained on the deviation parameters are of the same order of magnitude as those extracted from on-shell $W$ production, using the same angular variables. It is expected, however, that a more sophisticated analysis, using the method proposed in reference [20] could substantially increase the sensitivity to the trilinear boson couplings. Such an analysis seems to be suitable for the four-fermion production and it is important to be carried out, taking into account the precise experimental conditions.

In summary, we have presented a full tree-order calculation of the four-fermion production processes
$$e^+e^- \to \ell\bar{\nu}_\ell\, q\bar{q}'.$$
We have shown that a consistent treatment of the width effects is possible, at least for the fermionic contribution, and we have implemented it in a Monte Carlo programme. The C and P preserving deviations from the trilinear Standard Model couplings have been taken into account, and a first analysis on the sensitivity of the four-fermion production on these deviation parameters is presented.

### Acknowledgements

It is pleasure to thank R.Kleiss, D.Zeppenfeld, G.Gounaris, E.Argyres, W.J.Stirling, G.J. van Oldenborgh, R.Sekulin and S.Tzamarias for helpful discussions. This work is supported by the EU grant CHRX-CT93-0319.

# References

[1] K. Gaemers and G. Gounaris, Z.Phys. **C1**(1979) 259.

[2] M. Bilenky et al., Nucl. Phys. **B409** (1993) 22.

[3] K. Hagiwara, R. Peccei, D. Zeppenfeld and K. Hikasa, Nucl.Phys. **282** (1987) 253.

[4] E.N. Argyres and C.G. Papadopoulos, Phys. Lett. **B263** (1991) 298.

[5] C.G. Papadopoulos, Phys. Lett. **B333** (1994) 202.

[6] S.L.Glashow, Nucl.Phys. **22** (1961) 579;
S.Weinberg, Phys.Rev.Lett. **19** (1967) 1264;
A.Salam, *in Elementary particle theory*, ed. N.Svartholm (Almquist and Wiksell, Stockholm, 1968)p.367

[7] R. Kleiss and W.J. Stirling, Nucl. Phys. **B262** (1985) 235.

[8] M.Veltman, Physica **29** (1963) 161 and Physica **29** (1963) 186.

[9] F.A.Berends and G.B.West, Phys.Rev. **D1**(1970) 122;
A.Aeppli, F.Cuypers and G.J. van Oldenborgh, Phys.Lett. **B314** (1993) 413;
A.Aeppli, G.J. van Oldenborgh and D. Wyler, Nucl.Phys. **B428** (1994) 126;
M.Nowakowski and A.Pilaftsis, Z.Phys. **C 60** (1993) 121;
Y.Kurihara, D.Perret-Gallix and Y.Shimizu, '$e^+e^- \to e^-\bar{\nu}_e u\bar{d}$ from LEP to linear collider energies' KEK-PREPRINT-94-150,KEK-CP-021,LAPP-EXP-94-23, hep-ph:9412215.

[10] C.F. von Weiszäker, Z.Phys. **88** (1934) 612;
E.J.Williams, Phys.Rev. **45** (1934) 729;
M-S.Chen and P.Zerwas, Phys. Rev. **D12** (1975) 187.

[11] U.Baur, J.A.M.Vermaseren and D.Zeppenfeld, Nucl.Phys. **B375** (1992) 3.

[12] R.Kleiss and D.Zeppenfeld, private discussions.

[13] E.N.Argyres, W.Beenakker, A.Denner, S.Dittmaier, J.Hoogland, R.Kleiss, C.G.Papadopoulos, G.Passarino, and G.J. van Oldenborgh, in preparation.

[14] E.N.Argyres, G. Katsilieris, A.B. Lahanas, C.G.Papadopoulos and V.C. Spanos, Nucl.Phys. **B391** (1993) 23;
J. Papavassiliou and K. Philippides,Phys. Rev. **D 48** (1993) 4255.

[15] A. Denner, G. Weiglein and S. Dittmaier, 'Application of the Background-Field Method to the electroweak Standard Model', BI-TP.94/50, UWITP 94/03, October 1994.

[16] F.A.Berends, R.Pittau and R.Kleiss, Nucl.Phys.**B424**(1994) 308.

[17] R.Kleiss and R.Pittau, Comp.Phys.Commun. **83** (1994) 141.

[18] G.J. Gounaris and F.M. Renard, Z.Phys. **C59** (1993) 133.

[19] M.Kuroda, F.M.Renard and D.Schildknecht, Phys.Lett. **B183** (1987) 366
J.Maalampi, D.Schildknecht and K.H.Schwarzer, Phys.Lett. **B166** (1986) 361.

[20] R.L.Sekulin, Phys.Lett. **B338** (1994) 369.

[21] W.T. Eadie et al., *Statistical Methods in Experimental Physics*, North-Holland, Amsterdam, 1971.

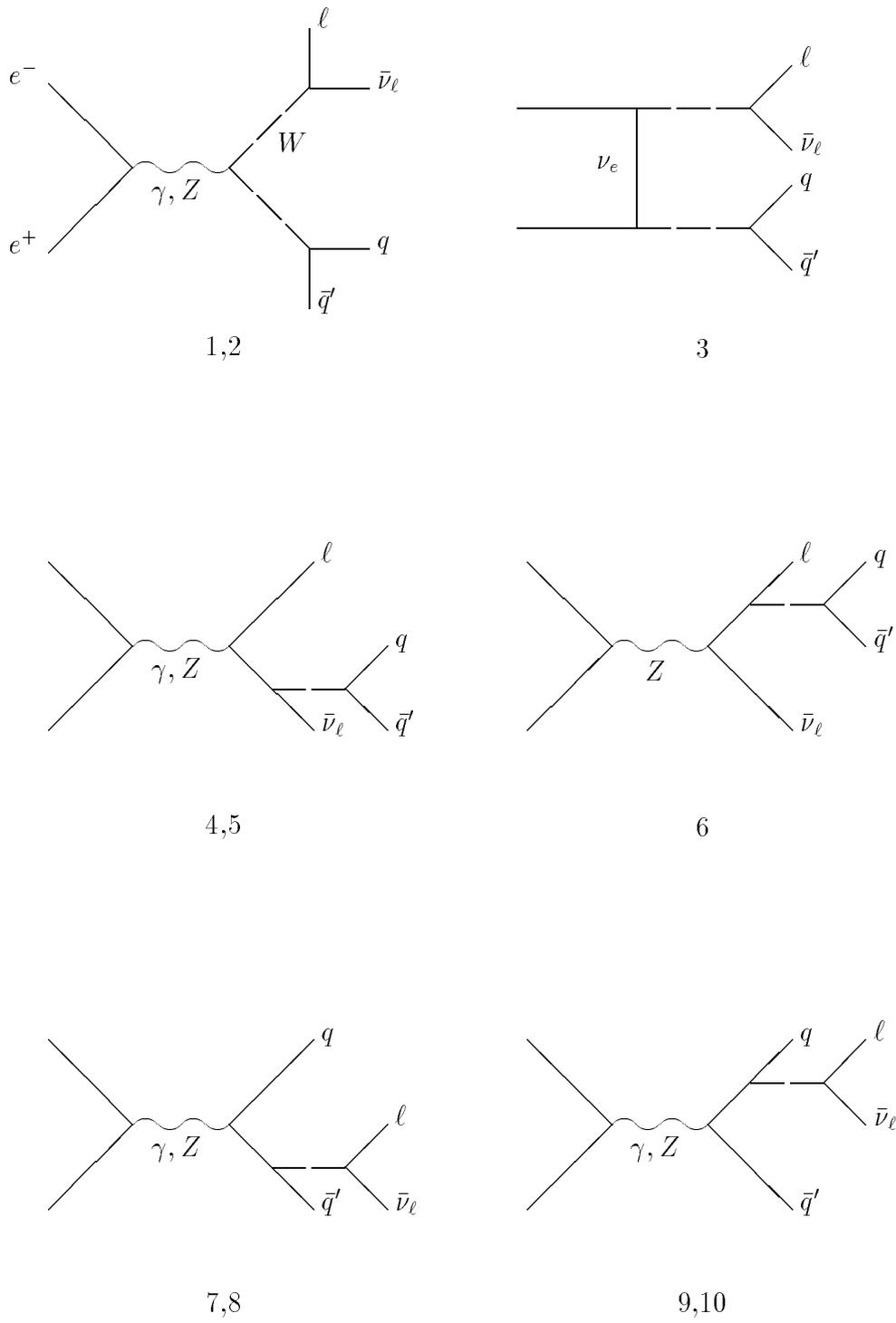

Figure 1: Feynman graphs contributing to $e^+e^- \to \ell\bar{\nu}_\ell\, q\bar{q}'$ in the s-channel.

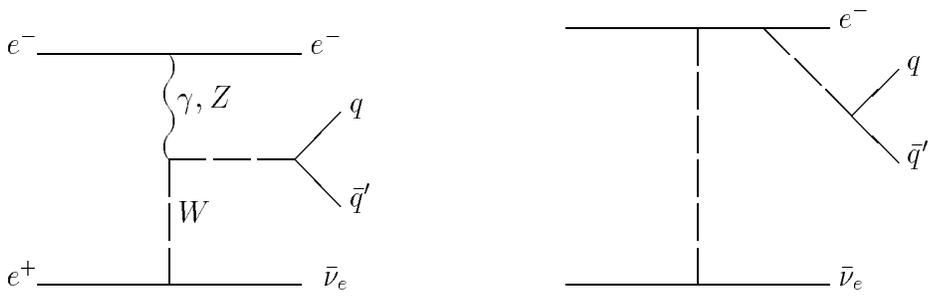
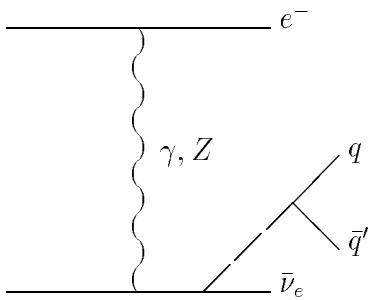
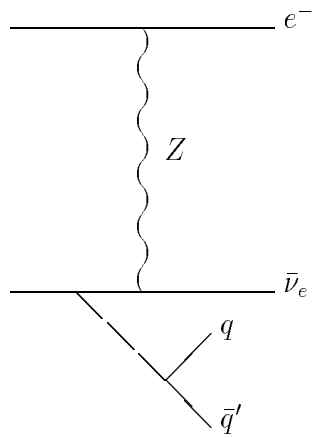
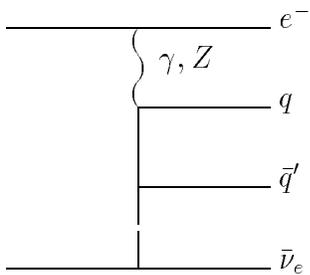
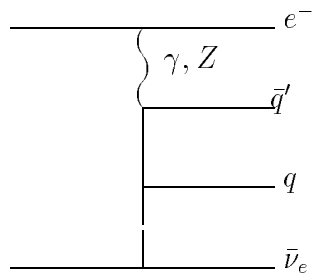

Figure 2: Feynman graphs contributing to $e^+e^- \to e\bar{\nu}_e\, q\bar{q}'$ in the t-channel.

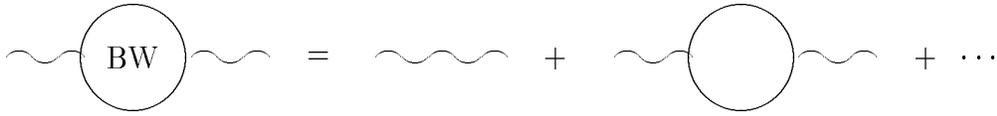

Figure 3: The Breit-Wigner propagator

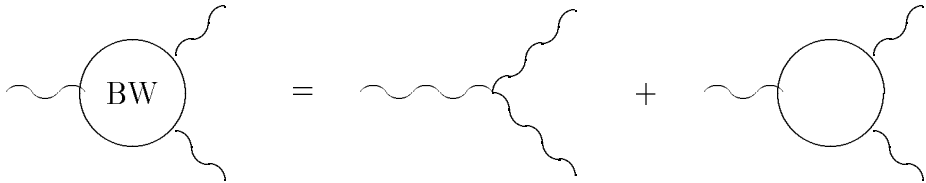

Figure 4: The Breit-Wigner triangle

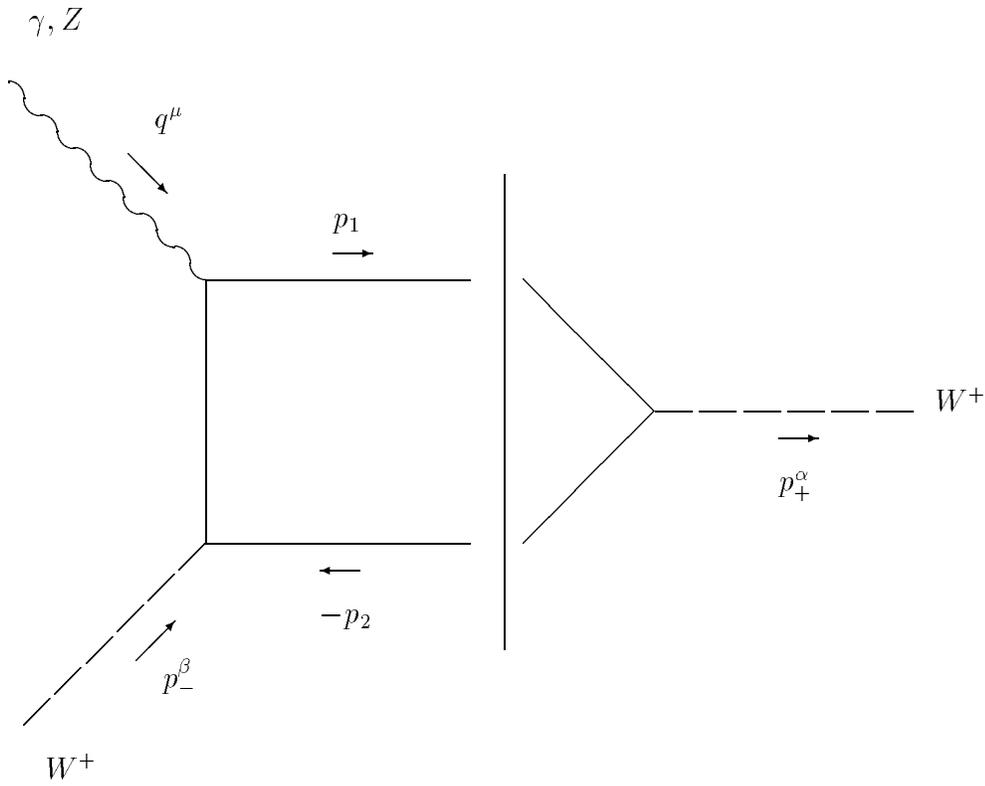

Figure 5: The cut graph contributing to the restoration of $U(1)$ gauge invariance.

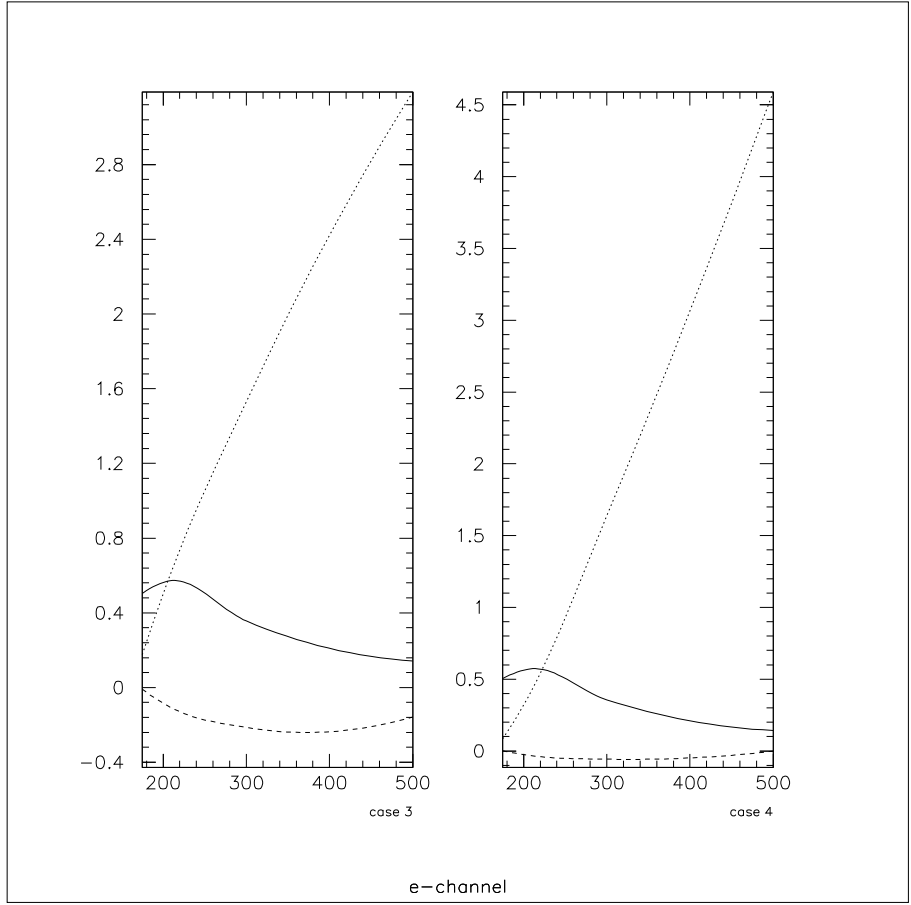

Figure 6: The total cross section, in picobarns, as a function of the centre-of-mass energy for cases 3 and 4 as explained in the text. The solid, dashed and dotted lines correspond to the Standard Model interference and signal terms respectively.

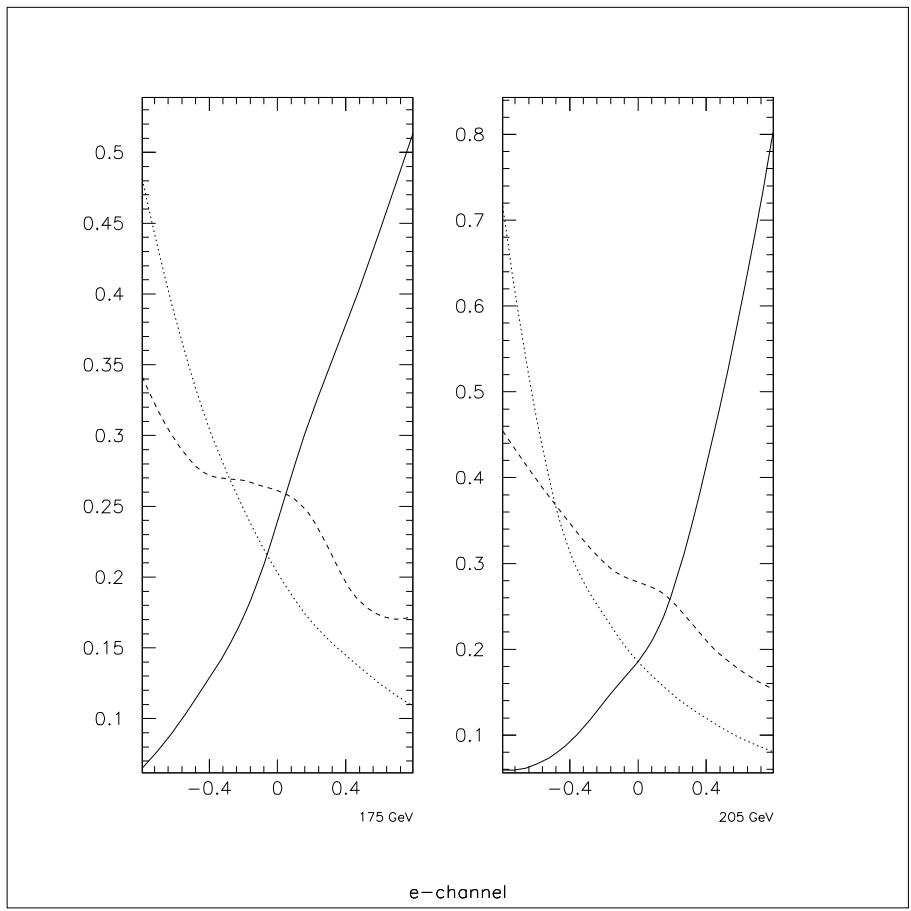

Figure 7: The Standard Model distributions $d\sigma/d\cos\theta_e$ (solid line), $d\sigma/d\cos\theta_W$ (dotted line) and $d\sigma/d\cos\theta_j^*$ (dashed line), in picobarns.

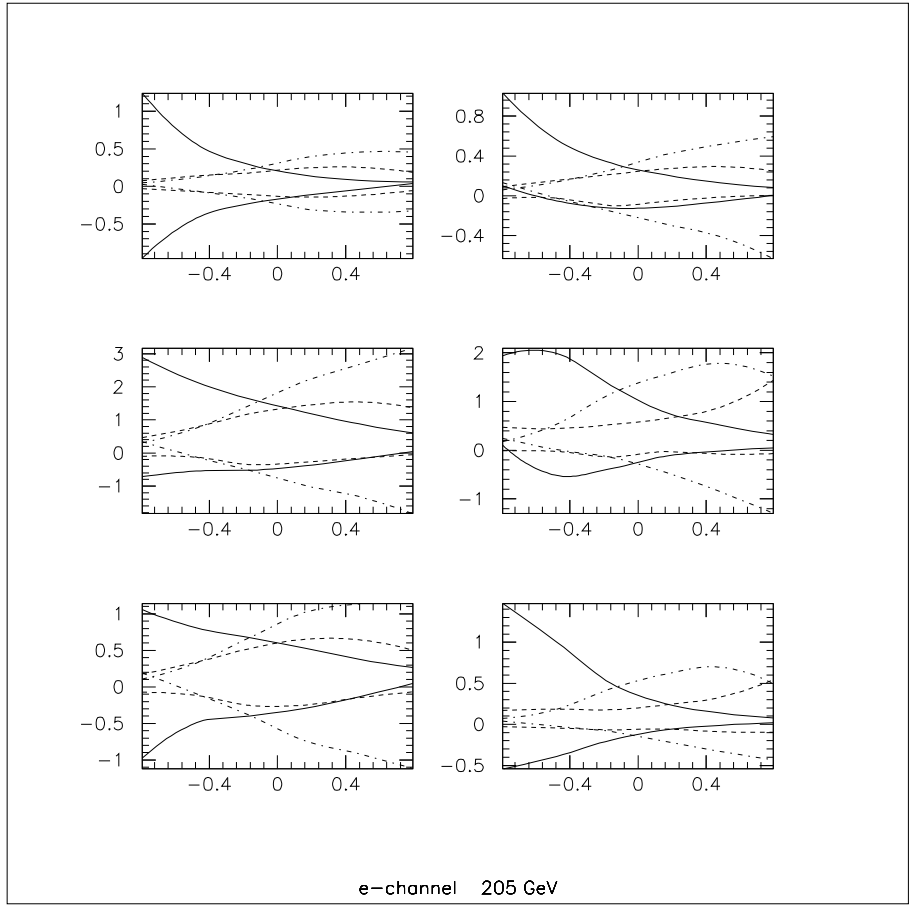

Figure 8: The ratio $\sigma^i/\sigma^0$ (lower lines) and $\sigma^s/\sigma^0$ (upper lines) for the six cases given in table 1 as a function of $\cos\theta_W$ (solid line), $\cos\theta_e$ (dot-dashed line) and $\cos\theta_j^*$ (dashed line).